\documentclass[10pt]{iopart}
\usepackage{graphicx}
\usepackage{amssymb}
\usepackage{iopams} 
\eqnobysec

\usepackage{tikz} 
\usepackage{pgfplots}
\usetikzlibrary{arrows.meta,arrows}
\usepackage[colorlinks,linkcolor=blue,urlcolor=black,citecolor=blue]{hyperref}

\newcommand{\beq}{\begin{equation}}
\newcommand{\eeq}{\end{equation}}
\newcommand{\be}{\begin{equation*}}
\newcommand{\ee}{\end{equation*}}
\newcommand{\beqa}{\begin{eqnarray}}
\newcommand{\eeqa}{\end{eqnarray}}
\newcommand{\bea}{\begin{eqnarray*}}
\newcommand{\eea}{\end{eqnarray*}}

\newcommand{\stackover}[2]{\mathrel{\mathop{#1}^{#2}}}
\newcommand{\mean}[1]{\langle#1\rangle}

\newcommand{\var}{\mathop{\rm Var}\nolimits}
\def\sign{\mathop{\rm sign}}
\newcommand{\dd}{{\rm d }}
\newcommand{\ii}{{\rm i}}

\newcommand{\sig}{{\sigma}}

\newcommand{\lam}{{\lambda}}
\renewcommand{\th}{{\theta}}
\newcommand{\gam}{{\gamma}}
\newcommand{\eff}{\mathrm{eff}}
\newcommand{\vsk}{\vskip4pt\noindent}

\begin{document}
\title{Interfaces of the two-dimensional voter model in the context of SLE}

\author{Claude Godr\`eche$^1$ and Marco Picco$^2$}

\address{$^1$ Universit\'e Paris-Saclay, CNRS, CEA, Institut de Physique Th\'eorique,
91191~Gif-sur-Yvette, France}
\address{$^2$ Sorbonne Universit\'e, Laboratoire de Physique Th\'eorique et Hautes \'Energies, CNRS UMR 7589, 4 Place Jussieu, 75005 Paris, France}

\begin{abstract}
This paper investigates various geometrical properties of interfaces of the two-dimensional voter model. 
Despite its simplicity, the model exhibits dual characteristics, resembling both a critical system with long-range correlations, while also showing a tendency towards order similar to the Ising-Glauber model at zero temperature. 
This duality is reflected in the geometrical properties of its interfaces, which are examined here from the perspective of Schramm-Loewner evolution.
Recent studies have delved into the geometrical properties of these interfaces within different lattice geometries and boundary conditions.
We revisit these findings, focusing on a system within a box of linear size $L$ with Dobrushin boundary conditions,
where values of the spins are fixed to either $+1$ or $-1$ on two distinct halves of the boundary,
in order to enforce the presence of a pinned interface with fixed endpoints (or chordal interface).
We also expand the study to compare the geometrical properties of the interfaces of the voter model with those of the critical Ising model and other related models.
Scaling arguments and numerical studies suggest that, while locally the chordal interface of the voter model has fractal dimension $d_{\rm f}=3/2$, corresponding to a parameter $\kappa=4$, it becomes straight at large scales, 
confirming a conjecture made by Holmes et al \cite{holmes}, and ruling out the possibility of describing the chordal interface of the voter model by SLE$_{\kappa}$, for any non zero value of $\kappa$.
This contrasts with the critical Ising model, which is described by SLE$_3$, and whose interface fluctuations remain of order $L$, and more generally with related critical models, which are in the same universality class.
\end{abstract}

\section{Introduction}
\label{sec:intro}

The purpose of this paper is to investigate the geometrical properties of the interfaces in the two-dimensional voter model \cite{clifford, holley}, a nonequilibrium system defined by simple dynamical rules.
In this model, an agent, located on a given site of a lattice, adopts the opinion (represented as a binary variable $\pm 1$) of a randomly selected neighbour.
Interfaces, in this context, separate regions of $+1$ and $-1$.

Like the Roman god Janus, the two-dimensional voter model has two faces looking in opposite directions.
By this metaphor is meant that within the same model coexist properties that are relatively contradictory.
On the one hand, the model has many characteristics of a critical system, in that it exhibits long-range correlations, the density of reactive interfaces (i.e., the fraction of $+-$ nearest neighbour pairs) decreases very slowly and the model can be defined as the zero-mass limit of the noisy voter model. 
On the other hand, the model demonstrates a tendency towards order, which is reminiscent of the Ising model evolving under Glauber dynamics at zero temperature, as the system eventually converges towards consensus%
\footnote{In two dimensions, a zero-temperature Ising system evolving under Glauber dynamics
may not always reach consensus (i.e., the ground state) but can be trapped into a metastable state characterised by stripes \cite{barros}.
The system nevertheless has an overall tendency towards order.}.
As we shall see, this duality of the properties of the model is reflected in the geometrical properties of its interfaces, which are the central focus of this study, approached here from the perspective of Schramm-Loewner evolution (SLE)\footnote{For an introduction to SLE for physicists, see \cite{cardy}.}.

We shall systematically compare these properties with those of the critical Ising model at equilibrium, which will serve as a backdrop in the present study.
Moreover, in order to introduce the notion of a `distance' between these two models---or even with the zero temperature Ising model---we shall use the fact that both the voter model and the Ising model with Glauber dynamics can be embedded in a broader class of two-dimensional nonequilibrium spin systems \cite{oliveira, drouffe}, including the Ising-Glauber model at any finite temperature, the noisy voter model and the majority vote model \cite{liggett,gray,oliv92,oliv23}.

Two recent papers, relevant for the present study, address the question of the geometrical properties of the interfaces of the voter model.
Reference \cite{tartaglia} examines the voter model on the square lattice in a box of linear size $L$ with periodic boundary conditions.
Over time, such a system reaches a consensus state where all opinions (or spins $\sigma=\pm1$) align. 
Consequently, the properties of interfaces can only be measured transiently,
within a temporal window, during which
the fractal dimension $d_{\rm f}$ of various bulk interfaces can be determined.
Considering the relation (discussed in section \ref{sec:fract}) between the fractal dimension $d_{\rm f}$ and the diffusion constant $\kappa$ defined in SLE, these
numerical measurements suggest a value of $\kappa=4$.
This prediction is in accord with the value of $\kappa$ obtained through
 the measurement of the variance of the winding angle (as detailed in section \ref{sec:fract}).

\begin{figure}[!ht]
\begin{center}
\includegraphics[width=5.6cm,height=5.6cm]{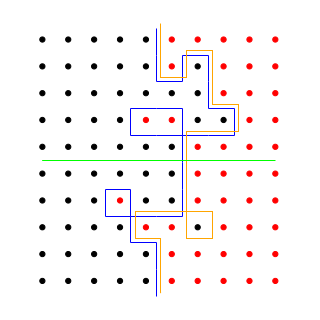}
\includegraphics[width=6cm,height=6cm]{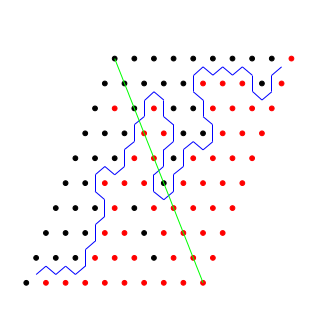}
\caption{\small
Square and triangular lattices with Dobrushin boundary conditions.
Positive spins are represented in black, negative spins in red.
Half of the spins on the boundary are fixed positive, the other half negative.
This enforces the existence of an interface with fixed endpoints, in blue.
For the square lattice, another choice of interface leads to the orange path.
See the text in section \ref{sec:MC} for more details.}
\label{fig:VL10}
\end{center}
\end{figure}

\begin{figure}[!ht]
\begin{center}
\includegraphics[width=6cm,height=6cm]{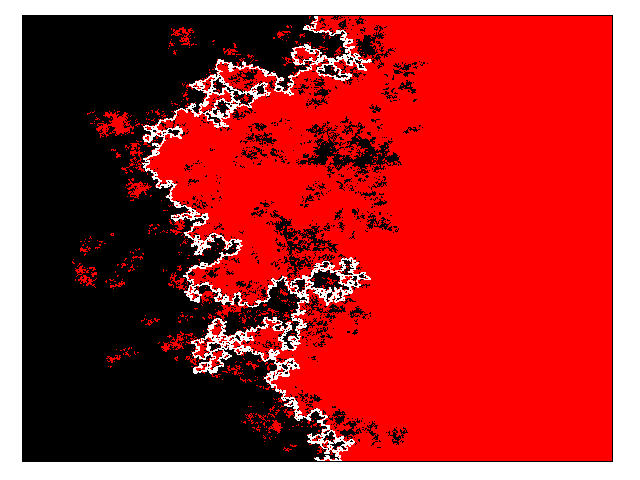}
\caption{\small
A configuration of the voter model at stationarity on the square lattice with Dobrushin boundary conditions ($L=1280$).
}
\label{fig:snapS}
\end{center}
\end{figure}

\begin{figure}[!ht]
\begin{center}
\includegraphics[width=9cm,height=6cm]{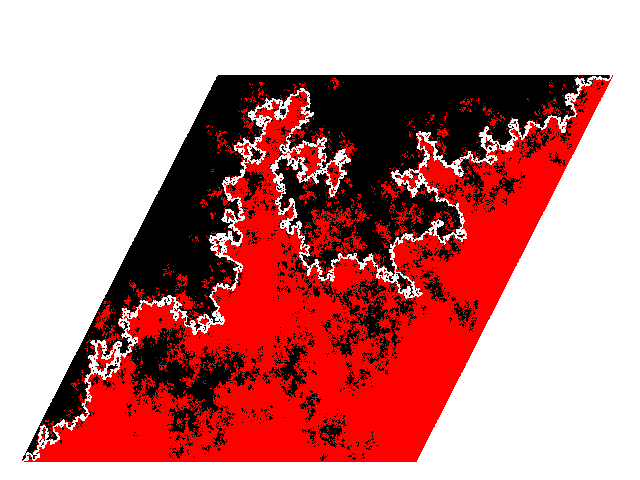}
\caption{\small
A configuration of the voter model at stationarity on the triangular lattice with Dobrushin boundary conditions ($L=1280$).
}
\label{fig:snapT}
\end{center}
\end{figure}

Another viewpoint is presented in \cite{holmes}, for the voter model on the triangular lattice, where the spins on the boundary of a box of linear size $L$ 
have fixed values, with half of the spins positive and the other half negative,
commonly referred to as 
\emph{Dobrushin boundary conditions} \cite{dobrushin,gallav} (see figure \ref{fig:VL10}).
These boundary conditions enforce the presence of an interface 
connecting two fixed endpoints located on the boundary of the box,
which we shall refer to as a pinned, or chordal interface.
In such a geometry the system becomes stationary, which avoids all the subtleties of having to deal with a transient range of time scales.
Figures \ref{fig:snapS} and \ref{fig:snapT} depict examples of configurations at stationarity with such boundary conditions, respectively on the square lattice and on the triangular lattice.
In \cite{holmes} the conjecture is made that this interface becomes straight when $L\to\infty$, which means that its width should scale more slowly than $L$,
and would also point towards $\kappa=0$, which seems contradictory with the prediction $\kappa=4$ mentioned above.
The fractal dimension of the pinned interface is also measured in \cite{holmes} and leads to a value compatible with $\kappa=4$, as in \cite{tartaglia}.


Hereafter, we shall revisit these questions and more generally investigate the geometrical properties of the interfaces in the voter model---both the pinned interface and interfaces in the bulk---in comparison with those of the critical Ising-Glauber model, as well as those of other models in the two-parameter family of models mentioned above.
We shall argue by scaling arguments and numerical studies that, on the one hand, the pinned interface of the voter model indeed becomes straight, in contrast with that of the Ising-Glauber model at criticality, and that, nevertheless, the value $\kappa=4$ holds at a more local scale.

The paper is structured as follows.
In section \ref{sec:family} we give details on the definition of the two-parameter family of models under study and highlight the singular status of the voter model in this family.
In section \ref{sec:MC} we provide methodological details on the numerical measurements.
Section \ref{sec:magn} demonstrates the influence of the boundary conditions on the magnetisation in the bulk, for both the voter model and the critical Ising model.
Sections \ref{sec:brok} to \ref{sec:loc} are devoted to the analysis of various geometrical characterisations of the interfaces on the square and triangular lattices for these two models.
In these sections, we shall focus primarily on two kinds of observables:
(i) local observables including the lengths of the interfaces (of clusters in the bulk and the pinned interface) and the winding angle of the pinned interface;
(ii) global observables pertaining to the location of the pinned interface, such as Schramm's left passage probability or the magnitude of its fluctuations.
We conclude with a discussion in section \ref{sec:discuss}.

\section{The voter and Ising-Glauber models as members of a two-parameter family of models}
\label{sec:family}

This section aims at highlighting the singular role played by the voter model in the two-parameter family of models whose overview follows, and to compare this model with the Ising-Glauber model.
This is essentially a self-contained reminder of \cite{drouffe} (see also \cite{oliveira}).
Here, the models are defined on the square lattice.
Generalising to the triangular lattice would require the introduction of additional parameters, with no conceptual changes.

We describe this class of models using the language of opinion dynamics \cite{castellano}.
Agents, located on the sites $i=1,2,\dots$ of a $L\times L$ square lattice, possess at a given time $t$ an opinion (they vote right or left, Republicans or Democrats), or a strategy (they buy or they sell), represented by the variables $\sigma_i=\pm1$, hereafter also referred to as the `spin' at site $i$.
At each time step, one of these agents is randomly selected and changes opinion, or strategy, i.e., $\sig\to-\sig$, with a probability per unit time $w(\sig)$---or flipping rate---depending on the opinions or strategies of his four neighbours located south, east, north, and west.

More precisely, $w(\sig)$ is a function of the 
`local field' $h=\sum_j\sig_j$, where the sum is taken over the index $j$ of the neighbours
of the flipping spin $\sig$.
This local field takes the values $\pm4,\pm2,0$.
We also enforce up-down symmetry, that is to say, 
\beq\label{eq:sym0}
w(\sig)\vert_h=w(-\sig)\vert_{-h}.
\eeq
This equality means that the tendency for a spin to flip (or change opinion) is the same if both the sign of the spin $\sig$ and that of its local field are reversed.
For example, the probability of opinion change for a Republican surrounded by three Democrats (and a Republican) is the same as that of a Democrat surrounded by three Republicans (and a Democrat).
Let us start with the voter and Ising-Glauber models.

\subsection{The flipping rates of the voter and Ising-Glauber models}

\vsk
\textit{Voter model.}
The selected agent changes opinion by choosing the opinion of one of his neighbours taken at random.
Thus, if $\sig=+1$, this spin flips (i.e., the agent changes opinion) with probability per unit time
\beq\label{eq:votertaux}
w(+)|_{h=4}=0,
\quad
w(+)|_{h=2}=\frac{1}{4},
\quad
w(+)|_{h=0}=\frac{1}{2},
\eeq
hence
\beq\label{eq:votertaux0}
w(+)|_{h=-2}=\frac{3}{4}, \qquad w(+)|_{h=-4}=1.
\eeq
In other words, $w(\sig)$ is equal to the fraction of disagreeing neighbours.
The expressions of the rates (\ref{eq:votertaux}) and (\ref{eq:votertaux0}) are encapsulated in
the expression\footnote{The generalisation to the case of the triangular lattice is
\be
w(\sig)=\frac{1}{2}\left(1-\frac{\sig h}{6}\right).
\ee
}
\be
w(\sig)=\frac{1}{2}\left(1-\frac{\sig h}{4}\right).
\ee

\vsk
\textit{Ising-Glauber model.}
The expression of the flipping rate 
reads
\beq\label{eq:glau}
w(\sig)=\frac{1}{2}(1-\sig\tanh(\beta\,h)),
\eeq
where $\beta=1/T$ is the inverse temperature.
So the values taken by the flipping rate read
\beqa\label{eq:wh}
w(\sig) |_{h=0}=\frac{1}{2},
\nonumber\\
w(\sig) |_{h=\pm2}=\frac{1}{2}(1\mp\sig\gamma),
\nonumber\\
w(\sig) |_{h=\pm4}=\frac{1}{2}\left(1\mp\sig\frac{2\gamma}{1+\gamma^2}\right),
\eeqa
where the parameter $\gam\in(0,1)$ is defined as \cite{glauber}
\beq\label{eq:gamdef}
\gam=\tanh 2\beta.
\eeq
In particular, at criticality, where 
$\gam_c=1/\sqrt{2}$,
if $\sig=+1$, this spin flips with rate
\bea
w(+)|_{h=4}=\frac{1}{2}-\frac{\sqrt{2}}{3}\approx 0.029,
\quad
w(+)|_{h=2}=\frac{1}{2}-\frac{1}{2\sqrt{2}}\approx 0.146,
\nonumber\\
w(+)|_{h=0}=\frac{1}{2},
\eea
hence
\be
w(+)|_{h=-2}=\frac{1}{2}+\frac{1}{2\sqrt{2}}\approx0.854, 
\qquad w(+)|_{h=-4}=\frac{1}{2}+\frac{\sqrt{2}}{3}\approx 0.971.
\ee
For the zero-temperature Ising-Glauber model,
opinion changes rely on a majority rule.
Thus, if $\sig=+1$, this spin flips with probability per unit time
\beq\label{eq:isingtaux}
w(+)|_{h=4}=0,
\quad
w(+)|_{h=2}=0,
\quad
w(+)|_{h=0}=\frac{1}{2},
\eeq
hence
\beq\label{eq:isingtaux0}
w(+)|_{h=-2}=1, \qquad w(+)|_{h=-4}=1.
\eeq
The expressions of the rates (\ref{eq:isingtaux}) and (\ref{eq:isingtaux0}) are encapsulated in
the expression
\beq\label{eq:ising0}
w(\sig)=\frac{1}{2}\left(1-\sig\sign h\right).
\eeq

\begin{figure}
\begin{center}
\begin{tikzpicture}[baseline=(current bounding box.center), very thick, scale=10.,domain=0:1]
\draw (0.,1.05) node {$\gamma_{\rm bulk}$};
\draw(-0.06,1) node {$1$};
\draw(-0.06,0.5) node {$1/2$};
\draw (-0.06,-0.06) node {$0$};
\draw(0.5,-0.06) node {$1/2$};
\draw(0.66,0.32) node {\bf MV};
\draw(0.57,0.62) node {\bf IG};
\draw(0.48,1.04) node {\bf V};
\draw(0.33,0.45) node {\bf NV};
\draw (1.05,0.) node {$\gamma_{\rm int}$};
\draw (1.0,-0.06) node {$1$};
\draw (0,0) -- (1,0);
\draw (0,1) -- (1,1);
\draw (0,0) -- (0,1);
\draw (1,0) -- (1,1);
\draw[red] (0.5,1) to [bend right=5] (0.7071,0.7071);
\draw[black,fill=black] (0.5,1.) circle (0.06ex);
\draw[black,fill=black] (0.7071,0.7071) circle (0.06ex);
\draw[black,fill=black] (1.,1.) circle (0.06ex);
\draw[red] (0.7071,0.7071) to [bend right=5] (1,0.417953);
\draw [-{Latex[length=4mm]},black](0.65,1.) -- (0.7,1.);
\draw [-{Latex[length=4mm]},black](0.35,1.) -- (0.3,1.);
\draw [-{Latex[length=4mm]},red](0.602,0.84) -- (0.618,0.82);
\draw [-{Latex[length=4mm]},blue](0.492,0.84) -- (0.491,0.82);
 \draw[dashed,blue] plot (\x,\x) ;
\draw[dashed, domain=0:1, smooth, variable=\y, blue] plot ({\y/(1+\y*\y)}, {\y});
\draw[dashed, domain=0:1, smooth, variable=\y, blue] plot ({2 * \y/(1+\y*\y)}, {\y});
\node[inner sep=0pt] (PVoter3) at (0.2,1.3)
 {\includegraphics[width=.25\textwidth,height=.25\textwidth]{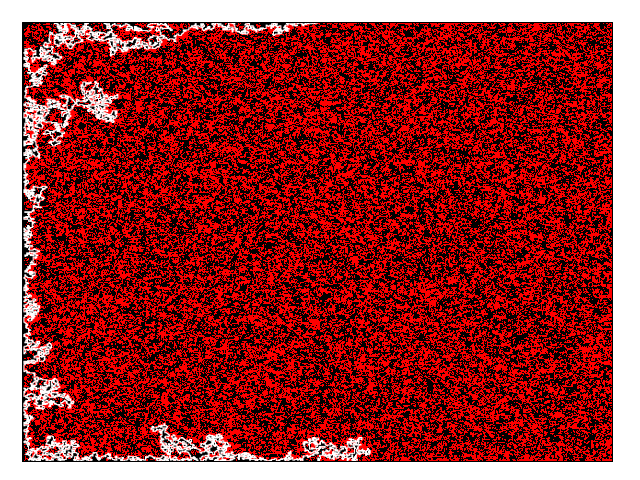}};
\node[inner sep=0pt] (Voter3) at (0.25,1)
{\includegraphics[width=.0025\textwidth]{PlotB1L640_3.png}};
\draw[<->,thick] (PVoter3.south) -- (Voter3.north west);
\node[midway] {$$};
\node[inner sep=0pt] (PVoter) at (0.8,1.3)
 {\includegraphics[width=.25\textwidth,height=.25\textwidth]{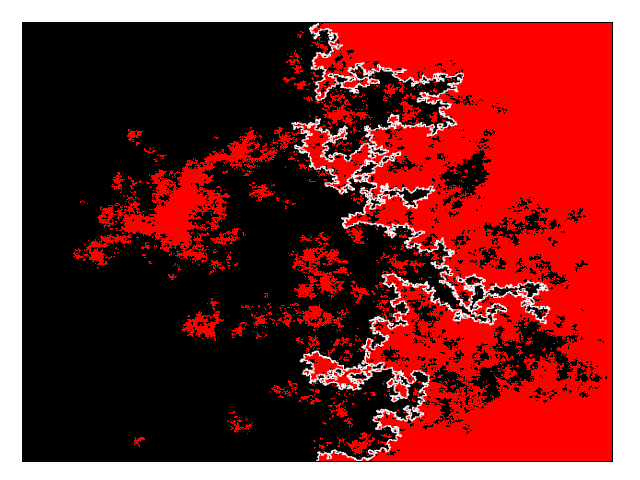}};
\node[inner sep=0pt] (Voter) at (0.5,1)
{\includegraphics[width=.0025\textwidth]{PlotB1L640.png}};
\draw[<->,thick] (PVoter.west) -- (Voter.north west)
 node[midway] {$$};
\node[inner sep=0pt] (PVoter2) at (1.2,1.1)
 {\includegraphics[width=.25\textwidth,height=.25\textwidth]{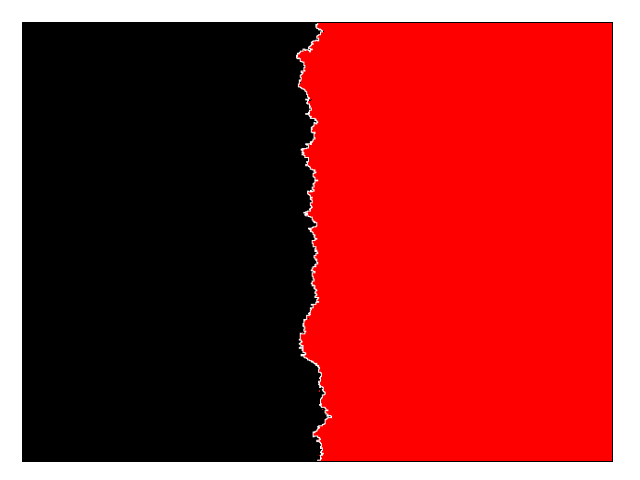}};
\node[inner sep=0pt] (Voter2) at (0.75,1.0)
{\includegraphics[width=.0025\textwidth]{PlotB1L640_2.png}};
\draw[<->,thick] (PVoter2.west) -- (Voter2.north west)
 node[midway] {$$};
\node[inner sep=0pt] (PCIM) at (1.2,0.7)
 {\includegraphics[width=.25\textwidth,height=.25\textwidth]{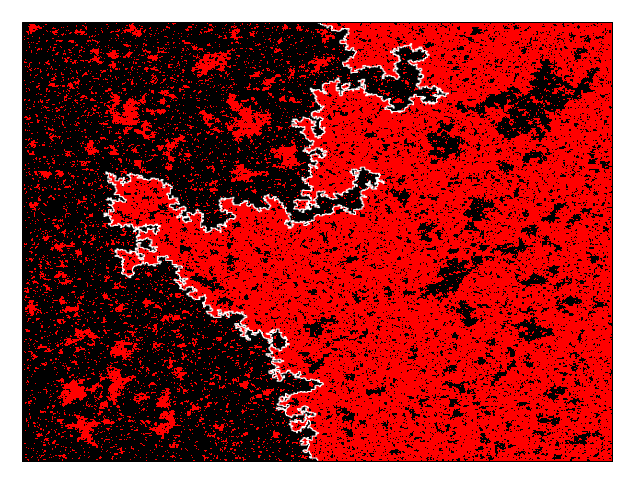}};
\node[inner sep=0pt] (CIM) at (0.7071,0.7071)
{\includegraphics[width=.0025\textwidth]{PlotCL640.png}};
\draw[<->,thick] (PCIM.west) -- (CIM.north west)
 node[midway] {$$};
\node[inner sep=0pt] (PNV) at (0.2,0.7)
 {\includegraphics[width=.25\textwidth,height=.25\textwidth]{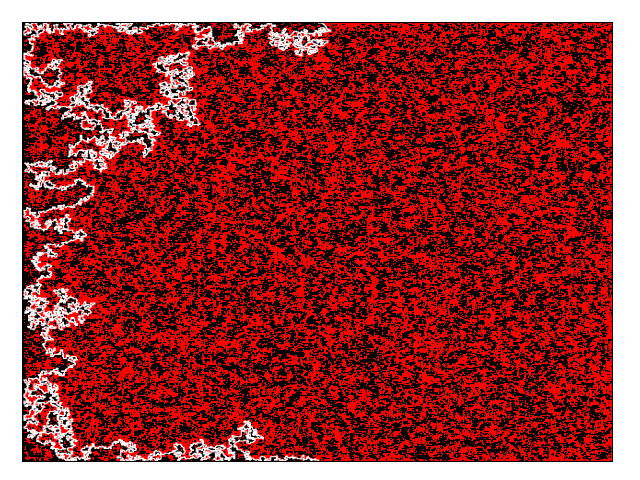}};
\node[inner sep=0pt] (NV) at (0.45,0.7071)
{\includegraphics[width=.0025\textwidth]{PlotNVL640.png}};
\draw[<->,thick] (PNV.east) -- (NV.north west)
 node[midway] {$$};
\end{tikzpicture}
\end{center}
\caption{Two-parameter family of models in the $(\gamma_{\rm int},\gamma_{\rm bulk})$ plane.
Dotted lines correspond respectively, from left to right, to the noisy voter model (NV) (see (\ref{eq:noisy})), the Ising-Glauber model (I-G) (see (\ref{eq:glaub})) and the majority vote model (MV) (see (\ref{eq:majo})).
The critical line (continuous) separates the high and low temperature regions.
The voter model corresponds to the point with coordinates ($\gamma_{\rm int}=1/2$, $\gamma_{\rm bulk}=1$).
It is unstable in four different directions depicted by arrows.
See the text for comments on the five realisations of stationary configurations, corresponding to a system size $L=640$.
}
\label{fig:phase}
\end{figure}

\subsection{Generalisation to a two-parameter family of models}
The flipping rate is defined as
\beq\label{eq:taux}
w(\sig)=\frac{1}{2}(1-\sig\tanh\left(\beta(h)\,h\right)),
\eeq
where $\beta(h)$ is some arbitrary function of $h$, obeying the symmetry constraint $\beta(-h)=\beta(h)$ inherited from (\ref{eq:sym0}), that is
\beq\label{eq:beta}
\beta(0)=0,\quad \beta(2)=\beta(-2),\quad \beta(4)=\beta(-4).
\eeq
Thus, the dynamics depends only on the two independent parameters 
 $\beta(2)$ and $\beta(4)$, hereafter denoted respectively by $\beta_{\rm int}$ and $\beta_{\rm bulk}$, where `int' stands for interfacial.
We also interpret $\beta_{\rm int}$ and $\beta_{\rm bulk}$ as the inverses of the temperatures
\be
T_{\rm int}=\frac{1}{\beta_{\rm int}}, \quad T_{\rm bulk}=\frac{1}{\beta_{\rm bulk}},
\ee
where the former corresponds to the motion of interfaces, and the latter to bulk excitations, as explained below.
In line with (\ref{eq:gamdef}), we finally introduce the parameters
\be
\gamma_{\rm int}=\tanh 2\beta_{\rm int},\qquad
\gamma_{\rm bulk}=\tanh 2\beta_{\rm bulk},
\ee
both defined in the range $(0,1)$.
So the values taken by the flipping rate read
\beqa\label{eq:wh}
w(\sig) |_{h=0}=\frac{1}{2},
\nonumber\\
w(\sig) |_{h=\pm2}=\frac{1}{2}(1\mp\sig\gamma_{\rm int}),
\nonumber\\
w(\sig) |_{h=\pm4}=\frac{1}{2}\left(1\mp\sig\frac{2\gamma_{\rm bulk}}{1+\gamma_{\rm bulk}^2}\right).
\eeqa

These two temperatures $T_{\rm int}$ and $T_{\rm bulk}$, or their inverses $\beta_{\rm int}$ and $\beta_{\rm bulk}$, or the parameters $\gam_{\rm int}$ and $\gam_{\rm bulk}$, 
are measures of the \emph{interfacial noise} and \emph{bulk noise} in the system, respectively.
This can be understood as follows \cite{drouffe}.
Consider the initial configuration where the system is divided by
a flat interface into two halves, one half with all spins
up, and the other one with all spins down, 
and periodic boundary conditions are assumed.
\begin{enumerate}
\item [$\ast$] If $T_{\rm int}=T_{\rm bulk}=0$ ($\gamma_{\rm int}=\gamma_{\rm bulk}=1$), 
implying 
\be
w(+) |_{h=4}=w(+) |_{h=2}=w(-) |_{h=-4}=w(-) |_{h=-2}=0,
\ee
this configuration will
not evolve in time, neither in the bulk, nor on the interface,
since all spins are surrounded by at least three spins of
the same sign. 
(This is the situation for the zero-temperature Ising-Glauber model.)

\item [$\ast$] However, if $T_{\rm bulk}=0$ ($\gamma_{\rm bulk}=1$) while $T_{\rm int}>0$ ($\gamma_{\rm int}<1$), implying $w(+) |_{h=4}=w(-) |_{h=-4}=0$, while $w(+) |_{h=2}$ and 
$w(-) |_{h=-2}$ are positive,
then spins at the interface will flip,
while those in the bulk will not.
(This is the situation for the voter model.)

\item [$\ast$] Conversely, if $T_{\rm int}=0$ ($\gamma_{\rm int}=1$) while $T_{\rm bulk}>0$ ($\gamma_{\rm bulk}<1$), then spins in the bulk will flip 
while those on the interface will not.
(They will later do so because of the noise coming from the bulk.)
\end{enumerate}

Note, however, that if the system initially consists of two parts separated by
a curved interface, for example a droplet of $+$ spins surrounded by a sea of $-$ spins, it will always evolve, even if $\gamma_{\rm int}=\gamma_{\rm bulk}=1$, since the local field of a given spin can be zero, and
$w(\sig) |_{h=0}=1/2$.

Let us illustrate these definitions by the examples of
the voter and Ising-Glauber models.
\vsk
\textit{Voter model.}
In the notations introduced above (see (\ref{eq:wh})), we have
\be
\gamma_{\rm bulk}=1\Longleftrightarrow T_{\rm bulk}=0,\quad 
\gamma_{\rm int}=\frac{1}{2} \Longleftrightarrow T_{\rm int}=\frac{2}{\rm arccoth\; 2}\approx 3.641.
\ee

\vsk
\textit{Critical Ising-Glauber model.}
This corresponds to the choice of parameters
\be
\gamma_{\rm bulk}= \gamma_{\rm int}=\frac{1}{\sqrt{2}} \Longleftrightarrow 
T_{\rm bulk}=T_{\rm int}=T_c=\frac{2}{\rm arccoth\; \sqrt{2}}\approx 2.269.
\ee

\vsk
\textit{Zero-temperature Ising-Glauber model.}
This corresponds to the choice of parameters
\be
\gamma_{\rm bulk}= \gamma_{\rm int}=1 \Longleftrightarrow T_{\rm bulk}=T_{\rm int}=0.
\ee
The dynamics is therefore only driven by the curvature of the interfaces between domains, which is a signature of coarsening \cite{bray}.

The voter model does not experience bulk noise, similar to the zero-temperature Ising-Glauber model, but is subject to interfacial noise, as the critical Ising-Glauber model.
Thus, in some sense, the voter model lies intermediate between the zero-temperature Ising-Glauber model and the critical Ising-Glauber model \cite{drouffe}.
This observation serves as the first indication supporting the claim that the voter model can be likened to a Janus model.

More generally, any point in the parameter space represented by any of the following choices of coordinates:
$(w(+)\vert_{h=2},w(+)\vert_{h=4})$, 
$(\gam_{\rm int},\gam_{\rm bulk})$,
$(T_{\rm int},T_{\rm bulk})$ or $(\beta_{\rm int},\beta_{\rm bulk})$,
represents a specific model of the class considered in the present work, as we now elaborate.

\vsk
\textit{Remark.}
On a spin operator basis, expression (\ref{eq:taux}) leads to the following form \cite{oliveira,cg2013}
\beqa\label{eq:wn}
\fl w(\sig)=\frac{1}{2}\left[1+\frac{1}{4}\Big(\gamma_{\rm int}-\frac{\gamma_{\rm bulk}}{1+\gamma_{\rm bulk}^2}\Big)
\sig(\sig_N\sig_E\sig_S+\sig_E\sig_S\sig_W+\sig_S\sig_W\sig_N+\sig_W\sig_N\sig_E)\right.
\nonumber\\
-\left.\frac{1}{4}\Big(\gamma_{\rm int}+\frac{\gamma_{\rm bulk}}{1+\gamma_{\rm bulk}^2}\Big)
\sig h\right],
\eeqa
where S, E, N, W stand for south, east, north, west.

\subsection{Remarkable lines in the two-parameter space}

There are several noteworthy lines in the two-parameter space of models.
The three first items of the list below were discussed earlier.
Two additional lines correspond to the noisy voter model and the majority vote model. 
Finally, we comment on the critical line.

\begin{enumerate}
\item Glauber dynamics corresponds to the line
\beq\label{eq:glaub}
\gam_{\rm int}=\gam_{\rm bulk}.\qquad (\mathrm{Ising-Glauber})
\eeq

\item The $\gam_{\rm bulk}=1$ line corresponds to models with no bulk noise ($T_{\rm bulk}=0$), hence the dynamics is only driven by interfacial noise.
Note that the effect due to the curvature of the interfaces is always present, as mentioned above.
\item The $\gam_{\rm int}=1$ line corresponds to models with no interfacial noise ($T_{\rm int}=0$), hence the dynamics is only driven by bulk noise.
In this case, the local spin aligns in the direction of the majority of its neighbours with probability one, if the local field $h=2$, i.e., if there is no consensus amongst the neighbours. If there is consensus amongst them, i.e., if $h=4$, the local spin aligns with its neighbours with a probability $<1$. 
Again, the effect due to the curvature of the interfaces is always present.
\item The noisy voter model corresponds to the choice
\be
w(\sig)=\frac{1}{2}\left(1-\frac{\lam}{4}\sig h\right),
\ee
with $0<\lam<1$. 
This corresponds to the line
\beq\label{eq:noisy}
\gam_{\rm int}=\frac{\gamma_{\rm bulk}}{1+\gam_{\rm bulk}^2}= \frac{\lam}{2}.
\qquad (\mathrm{noisy\ voter})
\eeq
Note that the four-spin operators disappear from the expression of the rate given by (\ref{eq:wn}).

\item For the majority vote model \cite{liggett,gray,oliv92,oliv23}, spins are aligned with the local field (i.e., with the majority of neighbours) with a given probability. 
More precisely, if $h\neq 0$
\beq\label{eq:mvm}
w(\sig)=\frac{1}{2}\left(1-\delta\, \sig\sign h\right)
\quad (0\le \delta\le1),
\eeq
and $w(\sig)=1/2$ if $h=0$. The model corresponds to the line
\beq\label{eq:majo}
\gam_{\rm int}=\frac{2\gam_{\rm bulk}}{1+\gamma_{\rm bulk}^2}=\delta,
\qquad (\mathrm{majority\ vote})
\eeq
when $\delta$ varies from 0 to 1,
i.e.,
\be
T_{\rm bulk}=2\,T_{\rm int}.
\ee
The rate (\ref{eq:mvm}) can be seen as a generalisation of the zero-temperature Ising rate (\ref{eq:ising0}).
\item The critical line separates the low-temperature phase (the upper-right corner) from the high-temperature phase (the rest of the square). 
This transition line was determined by finite-size scaling in \cite{oliveira} and by the following method in \cite{drouffe} (see also \cite{drouffe1}).
The distribution of the local mean magnetisation at a given site
$M_t=t^{-1}\int_0^t \dd u\,\sig(u)$ \cite{dornic}
is measured for large enough time. 
When crossing the critical point, the shape of the distribution changes from a broad profile with two maxima to a narrower one, with one maximum.
The outcome of these measurements is that, except for the voter model, interestingly 
located at the end of the critical line, all the models in the two-parameter space are in the Ising universality class, both for static properties \cite{oliveira} and for dynamical ones (such as features pertaining to persistence) \cite{drouffe}.

\end{enumerate}

To close, let us emphasise that apart along the Ising-Glauber line, none of these models satisfies detailed balance \cite{oliveira,cg2013}. 
At long times, along the half line $(\gam_{\rm int}\ge1/2,\gam_{\rm bulk}=1)$, a system with periodic boundary conditions goes to a consensus state where all opinions (or spins) align\footnote{Except at the zero-temperature Ising-Glauber point $(\gam_{\rm int}=1,\gam_{\rm bulk}=1)$, as mentioned previously.}. 
With the exception of this line, all other models reach a stationary state, whose measure is unknown, or an equilibrium state for the Ising-Glauber model, with Gibbsian measure.

\subsection{By way of summary}
Figure \ref{fig:phase} depicts the phase diagram of the two-parameter family of models under consideration. 
It contains three lines, shown in dashed blue, corresponding respectively to the noisy voter model (NV), the Ising-Glauber model (IG), and the majority vote model (MV). 
The critical line, separating the low-temperature phase (upper-right corner) from the high-temperature phase, is shown in red.

In this phase diagram, we are primarily interested in the behaviour of the voter model, with coordinates
($\gamma_{\rm int}=1/2$, $\gamma_{\rm bulk}=1$). 
This point is unstable, flowing away in four different directions.
In particular, along the critical line, the flow is toward the critical Ising model. 
The snapshots surrounding this phase diagram depict realisations of the stationary states reached by the system under Dobrushin boundary conditions in five situations, namely, going clockwise, for the voter model; for a generic point where $(1/2<\gamma_{\rm int}<1$, $\gamma_{\rm bulk}=1)$; for the critical Ising model; for the noisy voter model; and for a generic point where 
$(0<\gamma_{\rm int}<1/2$, $\gamma_{\rm bulk}=1)$.

\begin{figure}[!ht]
\begin{center}
\includegraphics[width=12cm,height=8cm]{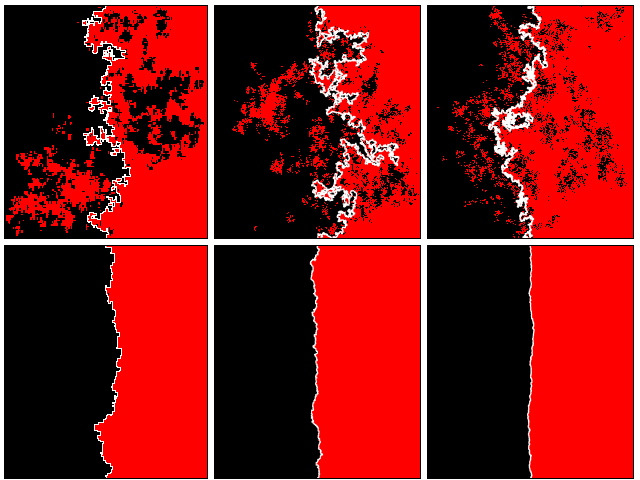}
\end{center}
\caption{Realisations of stationary configurations for the voter model ($\gamma_{\rm int}=1/2$, $\gamma_{\rm bulk}=1$) on top, and 
for the model with coordinates ($\gamma_{\rm int}=3/4$, $\gamma_{\rm bulk}=1$) on the bottom. 
From left to right, $L=160, 640$ and $2560$.} 
\label{fig:gameq1}
\end{figure}

\begin{figure}[!ht]
\begin{center}
\includegraphics[width=12cm,height=8cm]{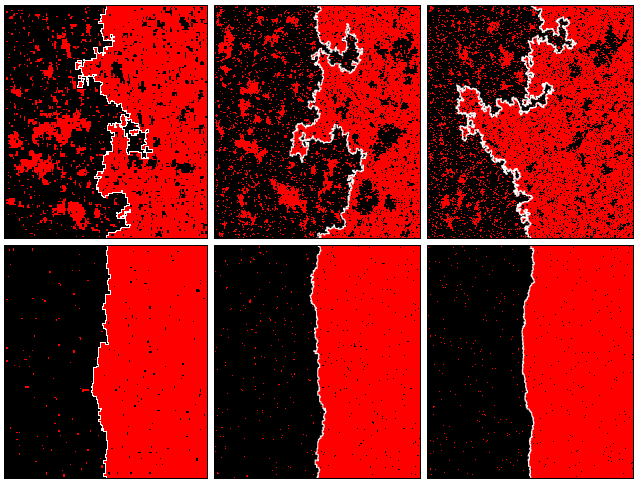}
\end{center}
\caption{Realisations of stationary configurations for the Ising-Glauber model at criticality on top, and for the model with coordinates $\gamma_{\rm int}=\gamma_{\rm bulk}=(1/\sqrt{2}+1)/2$ on the bottom. 
From left to right, $L=160, 640$ and $2560$.
} 
\label{fig:snapIM}
\end{figure}

The two high-temperature snapshots are representative of the disordered states attained by the system as soon as $\gam_{\rm int}<1/2$.
Both the snapshots at the voter point and at the critical Ising point exhibit fractal interfaces.
Quantifying the density of islands of the opposite colour on either side of the interface will be addressed when discussing, in section \ref{sec:magn}, the magnetisation profiles of these two models.
Finally, it is conspicuous on the snapshot for the generic point $(\gamma_{\rm int}=3/4,\gamma_{\rm bulk}=1)$ that the interface is much straighter.
In this regard, it is interesting to observe on figure \ref{fig:gameq1}
 that, while it is difficult to see the influence of the size of the system on the width of the interface for the voter model (top panels), for the generic point $(\gamma_{\rm int}=3/4,\gamma_{\rm bulk}=1)$, the interface becomes straighter when increasing the size $L$ (bottom panels).
 The same comments hold for figure \ref{fig:snapIM}.
A more quantitative description of these behaviours will be provided in \S~\ref{sec:loc}.

\section{Details of the Monte Carlo simulations}
\label{sec:MC}

In simulations, the initial condition is chosen either with all spins $+$ in the bulk, or completely disordered, while, 
on the boundary, half of the spins are fixed positive, the other half negative,
implementing Dobrushin boundary conditions, as in figure \ref{fig:VL10}.
The process is then iterated until the system reaches a stationary state. 
In practice, we calculate the autocorrelation time $\tau(L)$ associated with the interface length, and we start our measurements after $100\, \tau(L)$ updates, then take a measurement every $\tau(L)$ updates. 
If not specified otherwise, averages are made over $10^6$ independent configurations.

Coming back to the definition of the pinned interface, see figure \ref{fig:VL10}, we note that, while for the triangular lattice there is no ambiguity in defining the interface 
separating the spins connected to the left part from the spins connected to the right part (in blue),
for the square lattice however, there exists 
an ambiguity in the definition of the interface.
This occurs when going through a crossing between four spins with two $+$ and two $-$, with alternating values while going around the corner.
We are thus naturally led to define two interfaces. 
A first one such that the interface always closes around the $+$ spins, shown in blue in figure \ref{fig:VL10}. 
A second one such that the interface always closes around the $-$ spins, shown in orange.
For both lattices, we also draw a green line along which we locate the crossing position of the interface in order to measure certain quantities that will be detailed below.

\section{Magnetisation profiles}
\label{sec:magn}

A natural way to corroborate some of the features extracted from figures \ref{fig:gameq1} and \ref{fig:snapIM}, which depict realisations of stationary configurations for the voter model and the critical Ising model, as discussed in section \ref{sec:family}, is to consider their magnetisation profiles.
 In this section we restrict the study to the square lattice.
The magnetisation on the square $\mathbb{S}=\{(x,y)\in(1,L)\times(1,L)\}$ is defined as
\be
m(x,y)=\mean{\sig_{x,y}},
\ee
where the mean is taken over the stationary configurations.

\subsection{Voter model}

\begin{figure}[!ht]
\begin{center}
\includegraphics[angle=0,width=0.7\linewidth,clip=true]{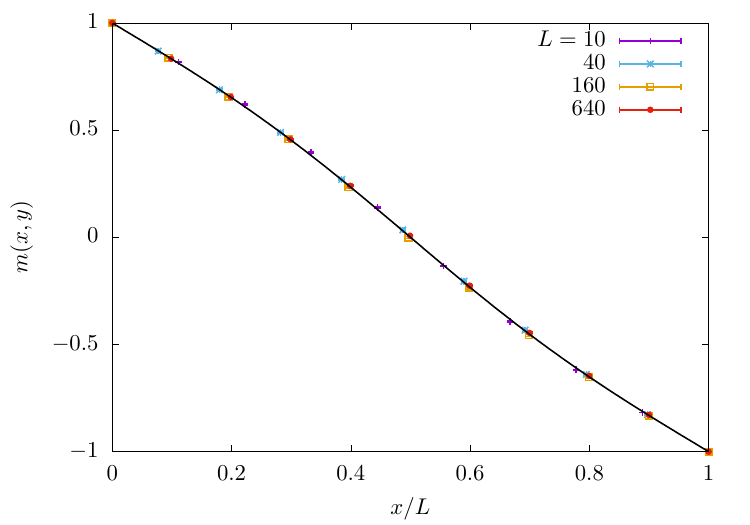}
\end{center}
\caption{Magnetisation profile $m(x,y)$ at $y=L/2$ (illustrated as a green line in figure \ref{fig:VL10}) for the voter model on the square lattice with Dobrushin boundary conditions against $x/L$.
Measurements are for $L=10,40,160,640$.
The black curve is the theoretical prediction (see (\ref{eq:resm}) and (\ref{eq:uvxy})).
}
\label{fig:magV}
\end{figure}

As observed in the snapshots for the voter model (see figure \ref{fig:gameq1}), big islands are present in the bulk, on both sides of the pinned interface. 
Since there is no bulk noise, these islands can only be produced by the pinned interface, and subsequently destroyed within the bulk or reabsorbed by the interface, resulting in a stationary situation with a non zero magnetisation profile $m(x,y)$.
Figure \ref{fig:magV} depicts this profile at mid-height, $y=L/2$, as measured in simulations, together with its theoretical prediction (\ref{eq:resm}) derived below, with perfect agreement.

The magnetisation of the voter model satisfies the Laplace equation $\Delta m=0$, where $\Delta$ is the discrete Laplacian on the lattice. 
We shall however use a continuum formalism for its determination, which,
as demonstrated by figure \ref{fig:magV}, is already accurate for small system sizes.
In the continuum, the theoretical expression for the magnetisation of the voter model, on the square $\mathbb{S}$
with Dobrushin boundary conditions, can be derived as follows. 
Initially, we compute this function in the upper half-plane $\mathbb{H}$, where $z=r\e^{\ii\th}$ 
($0\le\th\le\pi$), with boundary conditions $\sig=+1$ on the negative side of the horizontal axis, $\th=\pi$, and $\sig=-1$ on the positive side, $\th=0$.
We then proceed by applying the Schwarz-Christoffel mapping from $\mathbb{H}$ to the square $\mathbb{S}$.
In the half-plane $\mathbb{H}$, the magnetisation of the voter model is a harmonic function 
which solely depends on the polar angle $\th$, 
obeying $\Delta m(\th)=m''(\th)=0$, with boundary conditions $+1$ for $\th=\pi$ and $-1$ for $\th=0$.
This yields a linear function of the angle
\beq\label{eq:mth}
m(\th)=\frac{2\th}{\pi}-1.
\eeq
We then apply the Schwarz-Christoffel mapping $\mathcal{M}$ from the half-plane $\mathbb{H}$ to the square $\mathbb{S}$ to obtain $m(x,y)$,
\be
m(\th)\stackover{\longmapsto}{\mathcal{M}} m(x,y).
\ee
In practice, this goes as follows.
Define
\be
K\equiv K(k) =\int_{0}^{1} \frac{\dd u}{\sqrt{(1 - u^2)(1 - k^2 u^2)}}\approx 1.582,
\ee
which is the elliptic integral of the first kind, with parameter $k=(\sqrt{2}-1)^2$.
The mapping from $\mathbb{H}$ to the square $\mathbb{S}_K=\{(u,v)\in (-K,K)\times(0,2K)\}$, is obtained by 
\beq\label{eq:SC}
z(u,v)=\mathrm{sn}(w=u+\ii v,k),
\eeq
where $\mathrm{sn}$ is the Jacobi elliptic function.
From (\ref{eq:mth}), we infer that
\beq\label{eq:resm}
m(u,v)=\frac{2\arg z(u,v)}{\pi}-1.
\eeq
Finally, we replace $(u,v)$ by
\beq\label{eq:uvxy}
u=K\left(2\frac{x}{L}-1\right),\qquad v=2K \frac{y}{L},
\eeq
in order to obtain $m(x,y)$.

\begin{figure}[!ht]
\begin{center}
\includegraphics[angle=0,width=0.5\linewidth,clip=true]{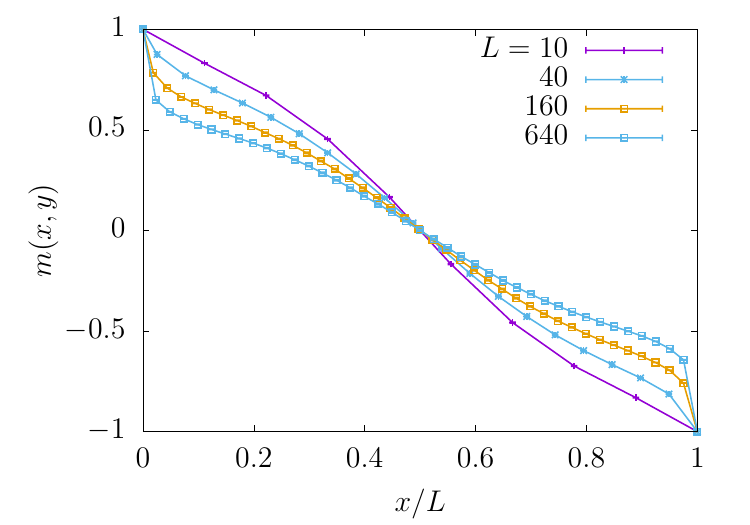}
\hskip -0.2cm
\includegraphics[angle=0,width=0.5\linewidth,clip=true]{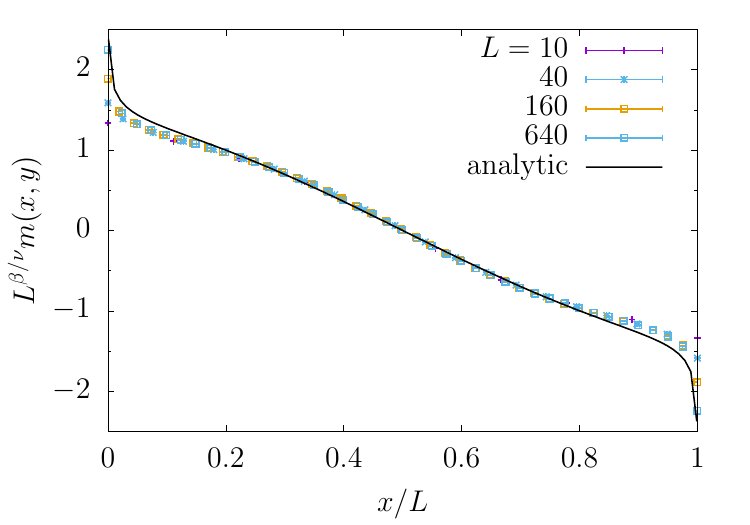}
\end{center}
\caption{Magnetisation profile for the critical Ising model on the square lattice with Dobrushin boundary conditions.
Left panel: $m(x,y)$ against $x/L$ for $y=L/2$ and for values of the system size $L$ given in the legend.
Right panel: $L^{\beta/\nu} m(x,L/2)$ against $x/L$ ($\beta/\nu=1/8$).
The black curve is the analytical expression for the infinite strip (see \cite{burk}).
}
\label{fig:magI}
\end{figure}

\subsection{Critical Ising model}
The situation is different for the critical Ising model since there is no magnetisation at criticality.
Hence, in general, for a finite system, the magnetisation should uniformly decrease to zero in the bulk as the system size $L$ becomes large. 
In the present situation, the magnetisation, which is finite close to the border, should decrease as a function of the distance from the border.
More precisely, in the bulk, it is expected to decrease as $L^{-\beta/\nu}$, in line with the scaling form \cite{fisher}
\beq\label{eq:fisher}
m(x,y) = \frac{1}{L^{\beta/\nu}} g\left(\frac{x}{L},\frac{y}{L}\right),
\eeq
where $\beta=1/8$ is the magnetic critical exponent and $\nu=1$ is the correlation length critical exponent, and
where the scaling function $g$ only depends on the boundary conditions.

This prediction is confirmed in figure \ref{fig:magI}. 
The left panel shows the magnetisation profile for $y=L/2$ as a function of $x/L$ for different system sizes $L$.
The right panel depicts the rescaled quantity $L^{\beta/\nu} m(x,L/2)$, demonstrating its convergence to a function of $x/L$, in accordance with (\ref{eq:fisher}).
It also depicts the analytical expression of 
the scaling function $g$ for the geometry of a strip $L \times M$ with $M\gg L$, and boundary conditions $+1$ on one side and $-1$ on the other side (black curve) (see equation (16) in \cite{burk}).
The agreement with the measured scaling function $g$ is surprisingly good.

A last comment is in order.
The snapshots of figure \ref{fig:snapIM} resemble those of the voter model in figure \ref{fig:gameq1}, though the bulk magnetisation of a critical Ising system vanishes in the thermodynamical limit.
The explanation is that the ratio of exponents $\beta/\nu$ being small, $L^{-\beta/\nu}$ remains appreciable even for a larger system size, consistently with what is observed in figure \ref{fig:magI}.

\section{Density of interfaces}
\label{sec:brok}

Let us start with the critical Ising model.
Consider the total length of bulk clusters---or number of broken bonds---discarding the contribution coming from the pinned interface.
This extensive quantity 
represents the energy of the system above the ground state.
Thus, in the thermodynamical limit, the rescaled total length converges to the equilibrium interface density of the critical Ising model
\be
\frac{\ell_{\mathrm{clusters}}(L)}{L^2}\to e=1-\frac{1}{\sqrt{2}},
\ee
for the square lattice, and
\be
\frac{\ell_{\mathrm{clusters}}(L)}{L^2}\to e=\frac{1}{2},
\ee
for the triangular lattice \cite{houtappel}.
Both predictions are very well confirmed by numerical simulations.
The contribution of the length of the pinned interface is subdominant, as discussed in section \ref{sec:fract}.

\begin{figure}[!ht]
\begin{center}
\includegraphics[angle=0,width=0.5\linewidth,clip=true]{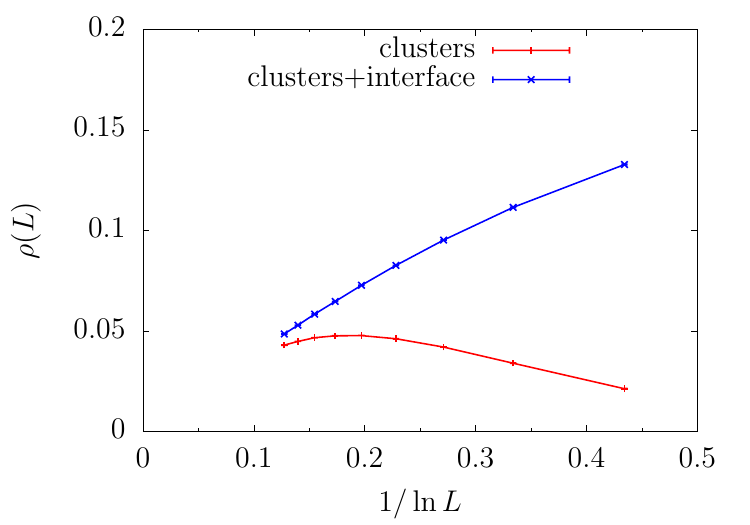}
\hskip -0.2cm
\includegraphics[angle=0,width=0.5\linewidth,clip=true]{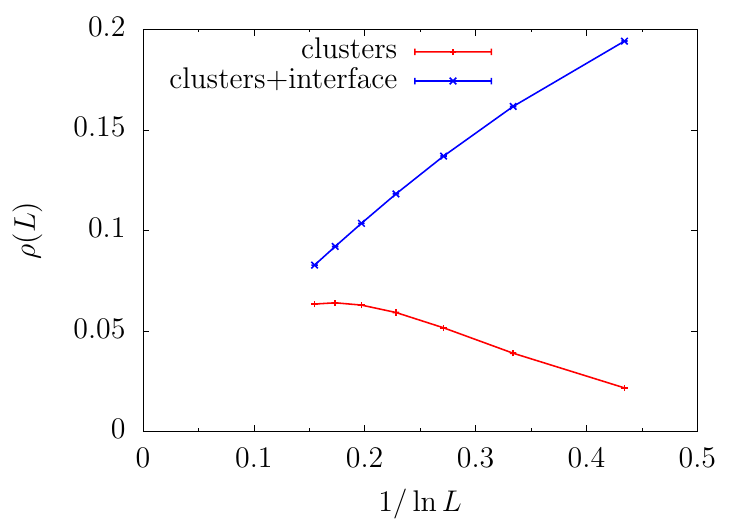}
\end{center}
\caption{Density of interfaces of the bulk clusters (in blue) and total density (in red) against $1/\ln{L}$ for the voter model on the square lattice (left) ($L=10,20,40,\dots,2560$), and on the triangular lattice (right)
($L=10,20,40,\dots,640$).
}
\label{fig:enV}
\end{figure}

We now turn to the voter model.
Let us recall that, on the square lattice with periodic boundary conditions, 
the density of broken bonds, that is to say,
 the fraction of neighbouring voters with opposite opinions (or reactive interfaces),
decays to zero as \cite{frachebourg,krb}\footnote{Hereafter,
$f(x)\simeq g(x)$ means that the two functions are asymptotically equivalent,
i.e., $f(x)/g(x)\to1$, when $x\to\infty$,
whereas the weaker form $f(x)\sim g(x)$
means that $f(x)/g(x)$ has much slower variations than $f(x)$ or $g(x)$ taken separately.}
\beq\label{eq:krap}
\rho(t)\simeq \frac{\pi}{2\ln t},
\eeq
reflecting the tendency of the process towards consensus.
A more precise calculation leads to\footnote{The original formula given in \cite{frachebourg} read $\rho(t)\simeq \pi/(2\ln t+\ln 256)$,
which is slightly incorrect.}
\be
\rho(t)\simeq \frac{\pi}{2\left(\ln(16 t)+\gam_{\mathrm{E}}\right)},
\ee
where $\gam_{\mathrm{E}}$ is Euler's constant.
In the present situation where the process reaches stationarity, if we trade time to space 
assuming diffusive scaling $t\sim L^2$, then this predicts
\be
\rho(L)\simeq \frac{\pi}{4 \ln L}.
\ee
For the triangular lattice the computation leads to
\beq\label{eq:krap}
\rho(t)\simeq \frac{\pi}{\sqrt{3}(\ln(24 t)+\gam_{\mathrm{E}})},
\eeq
hence the estimate
\be
\rho(L)\simeq \frac{\pi}{2\sqrt{3} \ln L}.
\ee
Figure \ref{fig:enV} depicts the density of interfaces $\rho(L)$ for the voter model on both the square and the triangular lattices, plotted against $1/\ln L$. 
The graph shows both the density of interfaces of the bulk clusters and the total density obtained by including the contribution from the pinned interface.
This demonstrates that, although the contribution from the pinned interface is subdominant (see section \ref{sec:fract}), its addition leads to improved convergence.
\section{Length of the pinned interface and fractal dimensions}
\label{sec:fract}

\subsection{Direct measurement of the fractal dimension}
\label{sec:direct}

\begin{figure}[!ht]
\begin{center}
\includegraphics[angle=0,width=0.7\linewidth,clip=true]{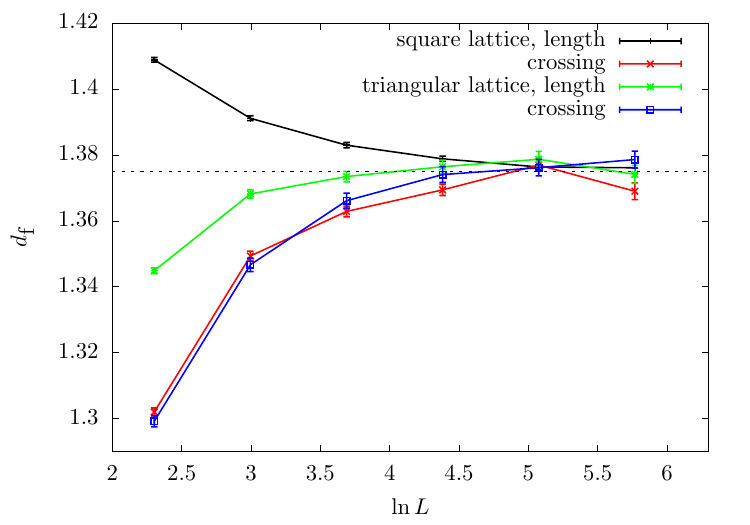}
\end{center}
\caption{Effective value of the fractal dimension $d_{\rm f}$ for the critical Ising model on the square and triangular lattices ($L=10,20,40,\dots,640$), obtained from (\ref{eq:length}) or from 
(\ref{eq:intersect}).
} 
\label{fig:dfI}
\end{figure}

The fractal dimension $d_{\rm f}$ of the pinned interface is obtained by measuring its length $\ell(L)$, then by using the fact that, for $L\gg1$, the latter scales as
\beq\label{eq:length}
\ell(L) \simeq L^{d_{\rm f}} .
\eeq
In practice, we shall measure the effective fractal dimension defined as
\beq\label{eq:dfeff}
d^{\eff}_{\rm f}(L) = \frac{\ln{\left({\ell(2 L)}/{\ell(L)} \right)}} {\ln{2}} .
\eeq
We start with the measurements of the fractal dimension of the pinned interface for the critical Ising model on both the square and triangular lattices, as depicted in figure \ref{fig:dfI}.
These measurements can be compared to the prediction of the continuum SLE$_\kappa$ theory,
which gives a relationship between $d_{\rm f}$ and the parameter $\kappa$ entering the theory, reading \cite{rhode,beffara}
\beq\label{eq:dfkappa}
d_{\rm f}=1+\frac{\kappa}{8}.
\eeq
For the critical Ising model, $\kappa=3$ \cite{chelkak}, entailing that $d_{\rm f}=11/8=1.375$, which is in accordance with the measurements shown in figure \ref{fig:dfI}.

By counting the number of times the interface crosses the green line in figure \ref{fig:VL10}
 at different scales $L$, one can gain an intuitive understanding of the fractal dimension of the interface. 
This is shown in table \ref{tab:cimcross}.
Additionally, one can extract the effective fractal dimension of this set of points from these data.
The dimension $d_{\times}$ of the intersection of the interface (of dimension $d_{\mathrm{f}}$) with the green line (of dimension $1$)
is given by the relation 
\beq\label{eq:intersect}
d_{\times}=1+d_{\rm f}-D,
\eeq
where $D=2$ is the embedding dimension \cite{falconer}.
Therefore adding unity to the effective dimension of the intersection yields another measure of 
$d_{\rm f}$, as shown in figure \ref{fig:dfI}.
 \begin{table}[h]
\caption{Mean number of times $\mathcal{N}_{\times}$ the interface crosses the green line in figure \ref{fig:VL10} for the critical Ising model on the square lattice (second line) and on the triangular lattice (third line).}
\label{tab:cimcross}
\begin{center}
\begin{tabular}{|c|c|c|c|c|c|c|c|}
\hline
$L$&10 & 20 & 40 & 80 & 160 & 320 & 640 \\
\hline
$\mathcal{N}_{\times}$ sq &1.19 & 1.46 & 1.87 & 2.40 & 3.10 & 4.02 & 5.21 \\
\hline
$\mathcal{N}_{\times}$ tri& 1.20 & 1.47 & 1.87 & 2.41 & 3.13 & 4.06 & 5.28 \\
\hline
\end{tabular}
\end{center}
\end{table}

\begin{table}[h]
\caption{Mean number of times $\mathcal{N}_{\times}$ the interface crosses the green line in figure \ref{fig:VL10} for the voter model on the square lattice (second line) and on the triangular lattice (third line).}
\label{tab:votercross}
\begin{center}
\begin{tabular}{|c|c|c|c|c|c|c|c|c|c|}
\hline
$L$&10 & 20 & 40 & 80 & 160 & 320 & 640 & 1280 & 2560 \\
\hline
$\mathcal{N}_{\times}$ sq&1.31 & 1.74 & 2.35 & 3.23 & 4.44 & 6.12 & 8.47 & 11.72 &16.30\\
\hline
$\mathcal{N}_{\times}$ tri& 1.32 & 1.74 & 2.36 & 3.23 & 4.44 & 6.13 & 8.47& & \\
\hline
\end{tabular}
\end{center}
\end{table}

\begin{figure}[!ht]
\begin{center}
\includegraphics[angle=0,width=0.7\linewidth,clip=true]{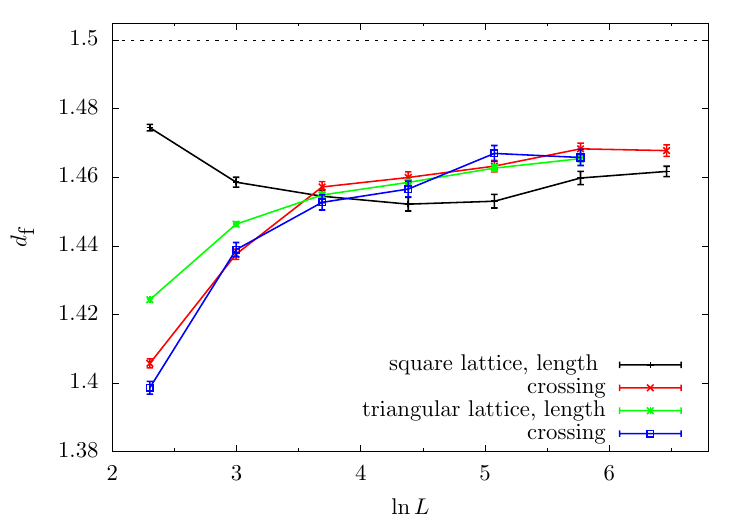}
\end{center}
\caption{Effective value of the fractal dimension $d_{\rm f}$ for the voter model on the square and triangular lattices ($L=10,20,40,\dots,640$), obtained from (\ref{eq:length}) or from (\ref{eq:intersect}).
} 
\label{fig:dfV}
\end{figure}

The same study is performed for the voter model in figure \ref{fig:dfV} and table \ref{tab:votercross}.
Figure \ref{fig:dfV} depicts the effective fractal dimension defined in (\ref{eq:dfeff}), for the voter model on the square and triangular lattices.
For the largest sizes attained, $d_{\rm f} \approx 1.465$ for both types of lattices.
Assuming that $d_{\rm f}$ converges to $3/2$ leads to the prediction $\kappa=4$.
Figure \ref{fig:dfV} also shows the measurement of the fractal dimension obtained from (\ref{eq:intersect}). 

Measurements of the fractal dimension $d_{\rm f}$ of the pinned interface of the voter model were made in \cite{holmes}, yielding similar results.

\subsection{Winding angle}

An alternative method for determining the fractal dimension of the pinned interface involves analysing the variance of its winding angle.
For an interface in a critical bidimensional system, 
the square of the variance of the difference of angles between the tangents of two points at a distance $d$ \cite{saleur,wieland} (see also \cite{schramm0}) is expected to satisfy the following relation
\be
\mean{\Theta^2(d)} \simeq {\kappa \over 2} \ln d,
\ee
where the parameter $\kappa$ is related to the fractal dimension $d_{\rm f}$ of the interface via (\ref{eq:dfkappa}).
Alternatively, one may consider the square of the variance of the difference of angles as a function of the length along the interface.
This length, denoted as $\ell(d)$, is expected to scale as 
$\ell(d)\simeq d^{d_{\rm f}}$ (see (\ref{eq:length})),
thus, using (\ref{eq:dfkappa}) (and keeping the same notation for the angle $\Theta$ as a function of $\ell$),
\beq\label{eq:w2}
\mean{\Theta^2(\ell)} \simeq {4 \kappa \over 8 + \kappa } \ln \ell.
\eeq
This approach is, from a numerical standpoint, more efficient.

\begin{figure}[!ht]
\begin{center}
\includegraphics[angle=0,width=0.7\linewidth,clip=true]{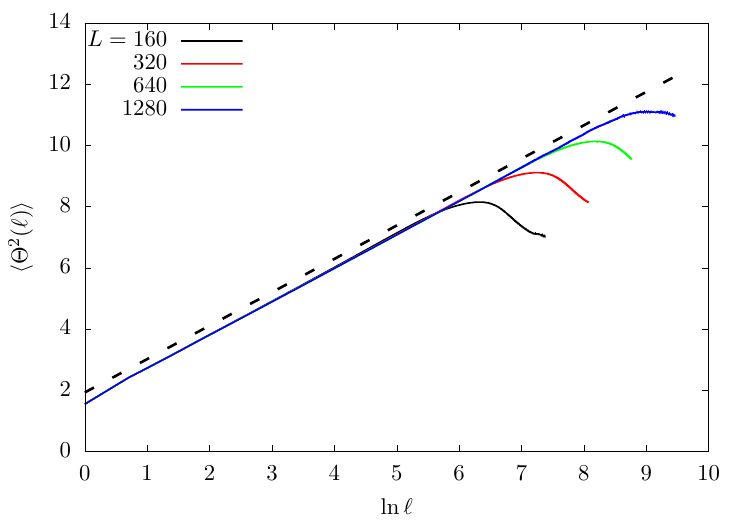}
\end{center}
\caption{Variance of the winding angle $ \mean{\Theta^2(\ell)}$ as a function of $\ln\ell$ 
where $\ell$ is the length along the pinned interface
of the critical Ising model on the square lattice with Dobrushin boundary conditions.
The dashed line corresponds to taking $\kappa=3$ in (\ref{eq:w2}) with arbitrary $y-$intercept, as a guide for the eye.
}
\label{fig:windI}
\end{figure}

In figure \ref{fig:windI}, we show measurements of $ \mean{\Theta^2(\ell)}$ for the critical Ising model on the square lattice.
The dashed line corresponds 
to taking $\kappa=3$ in (\ref{eq:w2}) with arbitrary $y-$intercept, as a guide for the eye, hence $d_{\rm f}=11/8$ by (\ref{eq:dfkappa}), which is quite convincing.

\begin{figure}[!ht]
\begin{center}
\includegraphics[angle=0,width=0.7\linewidth,clip=true]{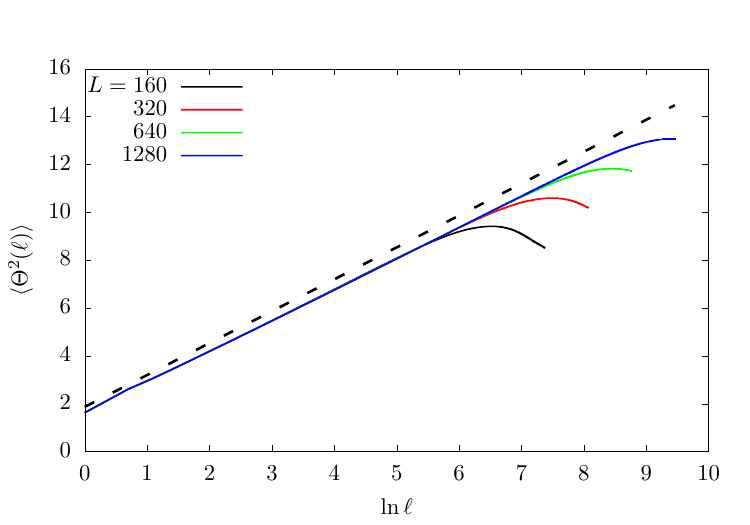}
\end{center}
\caption{Variance of the winding angle $ \mean{\Theta^2(\ell)}$ as a function of $\ln\ell$, where $\ell$ is the length along the pinned interface
 of the voter model on the square lattice with Dobrushin boundary conditions.
The dashed line corresponds to taking $\kappa=4$ in (\ref{eq:w2})
with arbitrary $y-$intercept, as a guide for the eye.
}
\label{fig:windV}
\end{figure}

The same analysis for the voter model on the square lattice leads to figure \ref{fig:windV}.
A best fit to (\ref{eq:w2}) yields $\kappa\approx 3.89$ for $L=1280$, which corresponds to $d_{\mathrm{f}}\approx 1.486$, slightly better than the value found in section \ref{sec:direct}.
The dashed line corresponds to $\kappa=4$ (as a guide for the eye), hence to $d_{\rm f}=3/2$, by (\ref{eq:dfkappa}).

\section{Schramm's left passage formula}
\label{sec:leftproba}

In light of the results from the previous section, and assuming SLE$_{\kappa}$ applies to the chordal interface of the voter model, one could be tempted to conclude that $\kappa=4$ for the latter.
However, a more global analysis, relying on Schramm's left passage probability formula \cite{schramm}, shows that 
the putative parameter $\kappa$ monotonically deviates from this value towards lower values with increasing system size,
ruling out the possibility of describing the chordal interface of the voter model by SLE$_{\kappa}$, for any non zero value of $\kappa$.

Consider again the upper half-plane $\mathbb{H}$.
Denote by $P_{\kappa}$ the probability that a curve connecting the boundary points at the origin and at infinity
passes to the left of a given interior point $z=r\e^{\ii\th}$. 
This probability depends only on the polar angle $\th$ of $z$.
The boundary conditions are $P_{\kappa}= 0$ when $\th = \pi$, and $P_{\kappa}=1$ when $\th=0$.
Schramm's formula reads
\beq\label{eq:Sleft1}
P_\kappa(\th) = \frac{1}{2} + \frac{\Gamma(4/\kappa)}{\sqrt{\pi}\Gamma((8-\kappa)/2\kappa)} {_2}F_1\left(\frac{1}{2}, \frac{4}{\kappa}, \frac{3}{2} ; -(\cot \th)^2\right) \cot \th,
\eeq
where $\Gamma$ is the Gamma function and ${_2}F_1$ is the hypergeometric function.
The mapping from $\mathbb{H}$ to the square $\mathbb{S}_K=\{(u,v)\in (-K,K)\times(0,2K)\}$ is then obtained by using the Schwarz-Christoffel mapping (\ref{eq:SC}) on (\ref{eq:Sleft1}).
Using the change of variables (\ref{eq:uvxy}) we finally obtain $P_{\kappa}(x,y)$ for the square 
$\mathbb{S}=\{(x,y)\in(1,L)\times(1,L)\}$.

In the present situation, using Schramm's formula for the pinned interface,
we can determine an effective value of the parameter $\kappa$ as follows.
Restricting to the square lattice,
we measure the left passage probability $P(x,y)$, with a numerical error denoted by 
$\Delta P(x,y)$, and compare it with the theoretical expression $P_{\kappa}(x,y)$ while varying $\kappa$. 
For a given size $L$, and for each value of $\kappa$, we compute \cite{chatelain}
\be
\chi(L,\kappa) = \frac{1}{L^2} \sum_{x,y}\left(\frac{P_\kappa(x,y) - P(x,y)}{\Delta P(x,y)}\right)^2,
\ee
which is minimised as a function of $\kappa$,
denoting the value of $\kappa$ for which $\chi_L(\kappa)$ is minimal by $\kappa_{\eff}(L)$.

\begin{figure}[!ht]
\begin{center}
\includegraphics[angle=0,width=0.5\linewidth,clip=true]{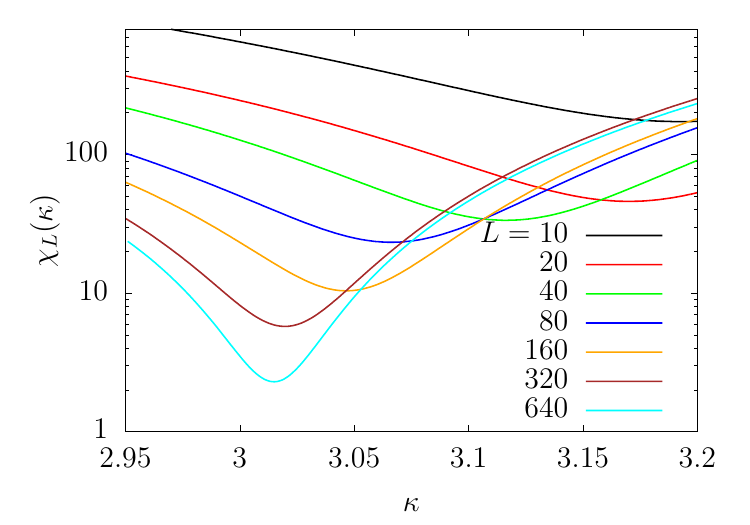}
\hskip -0.2cm 
\includegraphics[angle=0,width=0.5\linewidth,clip=true]{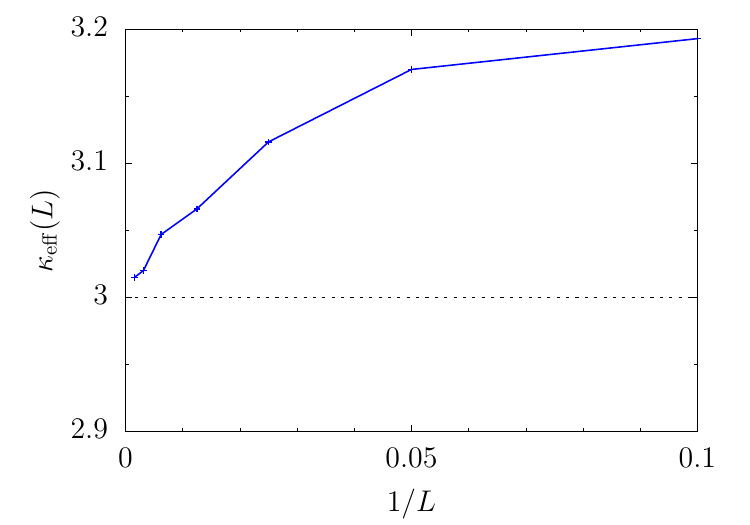}
\end{center}
\caption{Left panel: 
$\chi_L(\kappa)$ against $\kappa$ for the critical Ising model.
Right panel: $\kappa_{\mathrm{eff}}(L)$ against $1/L$.}
\label{fig:Mk}
\end{figure}
\begin{figure}[ht]
\begin{center}
\includegraphics[angle=0,width=0.5\linewidth,clip=true]{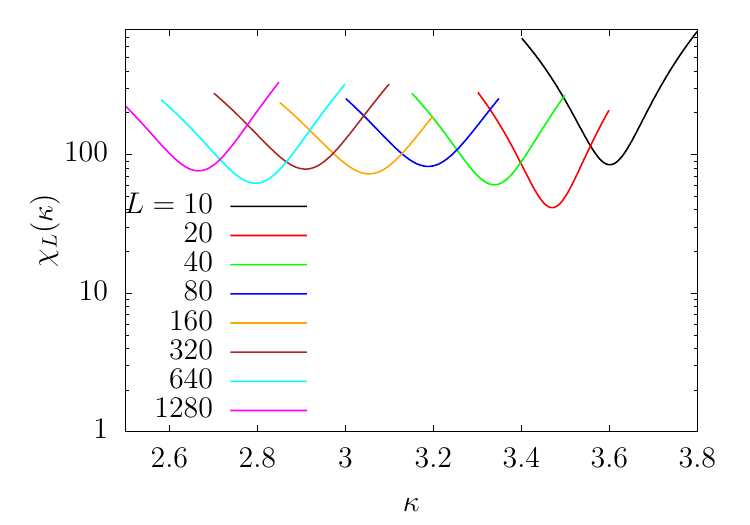}
\hskip -0.2cm 
\includegraphics[angle=0,width=0.5\linewidth,clip=true]{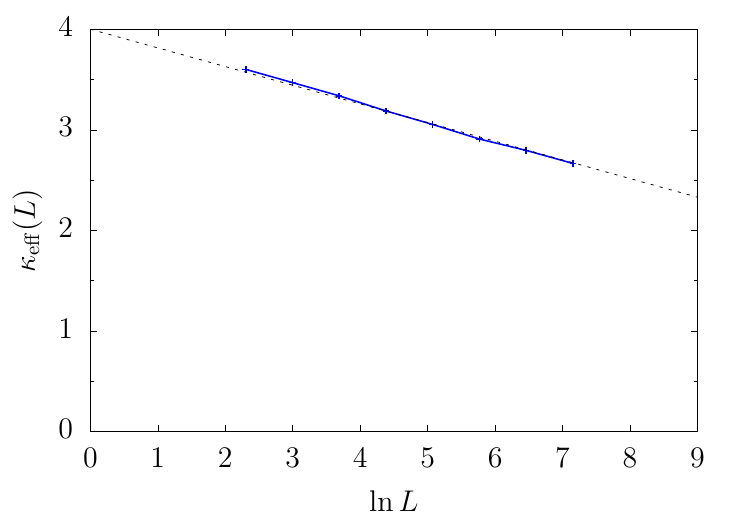}
\end{center}
\caption{Left panel: 
$\chi_L(\kappa)$ against $\kappa$ for the voter model.
Right panel: $\kappa_{\mathrm{eff}}(L)$ against $1/\ln L$.
The dotted line is a guide to the eye.
It was obtained as a best fit to the curve.}
\label{fig:kappa}
\end{figure}

We start with the analysis of the critical Ising model.
In the left panel of figure~\ref{fig:Mk}, we show the results of the measurements of $\chi_L(\kappa)$ as a function of $\kappa$ for various sizes. 
The sequence of curves drifts to the left while deepening,
i.e., $\chi_L(\kappa)$ decreases as a function of $L$. 
Since the number of independent configurations is constant ($10^6$ for each linear size), this decay indicates that the numerical measurements approach Schramm's formula as the system size increases.
The measured values of $\kappa_{\eff}(L)$ are entirely compatible with the theoretical prediction $\kappa=3$ \cite{chelkak} (see right panel of figure \ref{fig:Mk}).

Unlike the behaviour depicted in figure \ref{fig:Mk}, no convergence is observed as $L$ increases in the case of the voter model (see the left panel of figure \ref{fig:kappa}). 
The measured effective $\kappa_{\eff}(L)$ decreases linearly on a logarithmic scale, reaching a value of 2.65 when $L = 1280$. 
It is expected to continue decreasing for larger values of $L$ (see the right panel of figure \ref{fig:kappa}).
Moreover, the minimal value of $\chi_L(\kappa)$ does not decrease as a function of $L$, meaning that the measured probability
does not converge towards Schramm's formula as the system size $L$ increases.

As a conclusion, the left passage probability $P(x,y)$ measured for the voter model does not coincide with Schramm's formula for any finite value of $\kappa$,
entailing that the pinned interface is not conformally invariant under SLE. 
Assuming that, in the limit of large $L$, $\kappa_{\eff}(L)$ flows towards zero, this would imply, by (\ref{eq:dfkappa}), that formally $d_{\rm f}=1$, indicating that the pinned interface becomes straight, or, in other words, localises.
Sections \ref{sec:loc} is devoted to substantiating this observation.
As mentioned earlier, the straightening of the pinned interface has been previously conjectured in \cite{holmes}.

\section{Localisation of the interface}
\label{sec:loc}
\subsection{Crossing probability}

\begin{figure}[!ht]
\begin{center}
\includegraphics[angle=0,width=0.5\linewidth,clip=true]{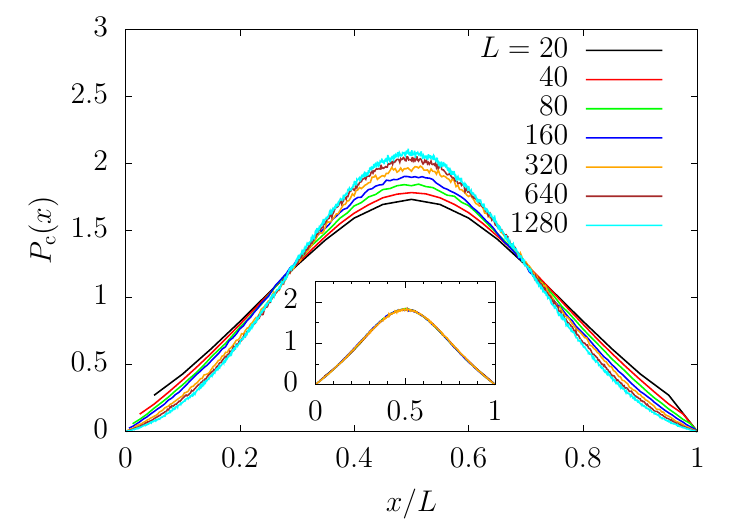}
\hskip -0.2cm
\includegraphics[angle=0,width=0.5\linewidth,clip=true]{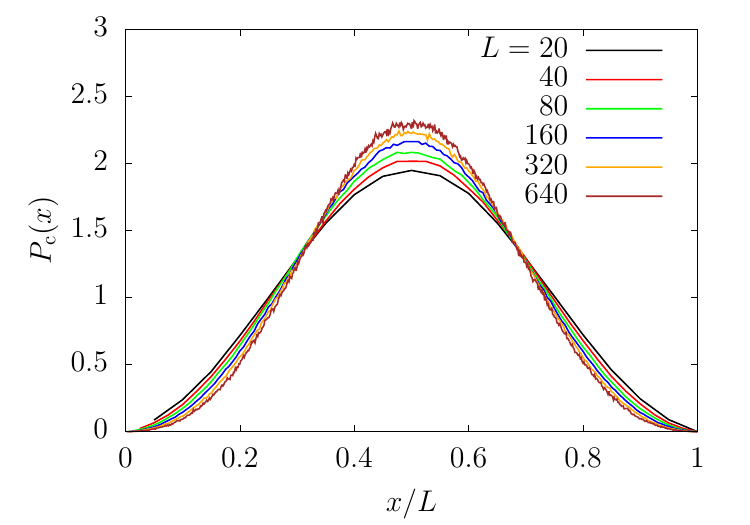}
\end{center}
\caption{Probabilities $P_{\mathrm{c}}(x)$ against $x/L$ for the voter model on the
square lattice (left panel) and the triangular lattice (right panel).
Measurements for the critical Ising model on the square lattice ($L=20, 40, 80, 160, 320$), are depicted in the inset of the left panel.
}
\label{fig:PC}
\end{figure}

A way to characterise the localisation properties of the chordal interface is by examining the probability 
$P_{\mathrm{c}}(x)$ that it crosses the green line in figure \ref{fig:VL10} at a specific point $x$.
Figure \ref{fig:PC} depicts this probability as a function of $x/L$ for the voter model. 
Measurements done on the square lattice are depicted in the left panel, while those done on the triangular lattice are shown in the right panel. 
In both cases, the crossing probabilities become increasingly centred as the system size increases. 
In contrast, measurements for the critical Ising model on the square lattice (for $L=20, 40, 80,160$ and $320$), shown in the inset of the left panel,
show no size dependency. 

\subsection{The width of the interface}

Moreover, we can quantitatively characterise the localisation properties of the chordal interface by measuring the magnitude of its fluctuations. 
To achieve this, we consider the following two quantities.

\begin{figure}[!ht]
\begin{center}
\includegraphics[angle=0,width=0.7\linewidth,clip=true]{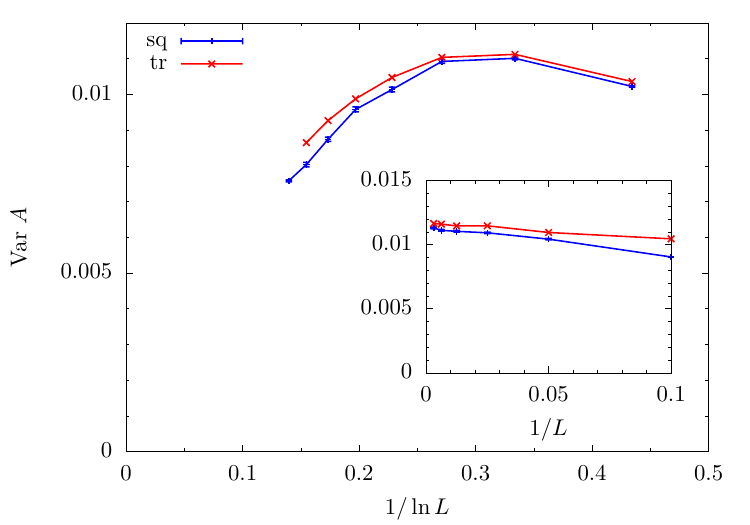}
\end{center}
\caption{Variance of the rescaled area to the left of the interface for the voter model plotted against $1/\ln{L}$ (up to $L=1280$).
In the inset, $\var A$ for the critical Ising model is plotted against $1/L$ 
(up to $L=320$).
}
\label{fig:area}
\end{figure}

Consider first the area to the left of the interface, which is calculated by counting all the sites on the left-hand side. 
This area, scaled by $L^2$ and denoted by $A$, fluctuates around $1/2$, so its variance 
$\var A=\mean{(A-1/2)^2}$, depicted in figure \ref{fig:area}, provides an insight into how the interface straightens.
While for the critical Ising model the variance converges towards a constant value (see the inset in figure \ref{fig:area}), for the voter model, on the contrary, this quantity decreases at larger values of $L$.
This decrease is slow, presumably logarithmic.
We therefore choose, in figure \ref{fig:area}, to plot the data against $1/\ln{L}$.

\begin{figure}[!ht]
\begin{center}
\includegraphics[angle=0,width=0.7\linewidth,clip=true]{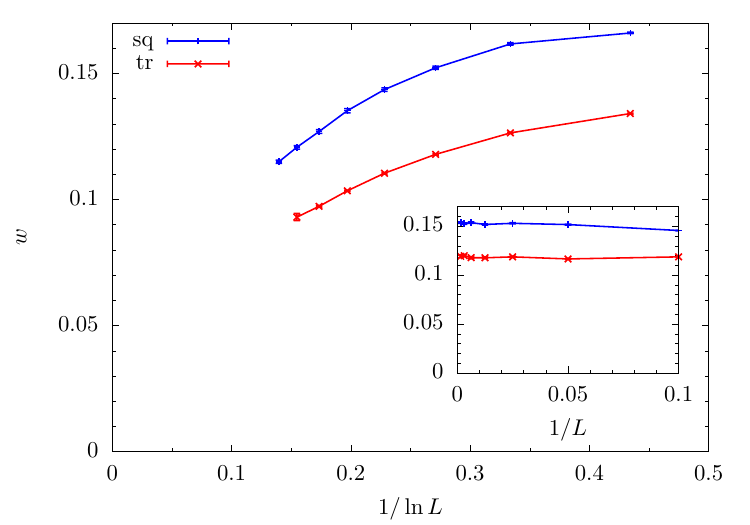}
\end{center}
\caption{Width $w$ against $1/\ln{L}$ for the voter model on the square and triangular lattices.
In the inset: same quantity against $1/L$ for the critical Ising model on the square and triangular lattices.
}
\label{fig:FigV}
\end{figure}

The second quantity we investigate is defined as follows.
Consider again the multiple crossings of the green line in figure \ref{fig:VL10} by the interface (see section \ref{sec:fract}).
The positions of these crossings, in reduced units, are denoted as $0\le \xi_i\le 1$ ($i=1, \dots, 2k +1$).
Define
\be
f = \xi_1 - \xi_2 + \xi_3 - \cdots,
\ee
which is the fraction of the interval $(0,1)$ that lies to the left of the interface (excluding islands of opposite sign). 
This is also the section by the green line of the area $A$ to the left of the interface.
Likewise, the section of the area $A$ to the right of the interface by the green line is equal to $1-f$.
Both quantities fluctuate around $1/2$.
Their fluctuations are given by
\be
w=\mean{(f-1/2)^2}.
\ee
This quantity is shown in figure \ref{fig:FigV} as a function of $1/\ln L$ for both the square and triangular lattices.
Its decay is, in all respects, similar in nature to that observed in figure \ref{fig:area} for the area to the left of the interface.
For comparison, we show this quantity for the critical Ising model in the inset.

\section{Discussion}
\label{sec:discuss}

In this article, we have gathered a convergent set of evidence supporting the conjecture put forth in \cite{holmes} regarding the straightening of the chordal interface in the two-dimensional voter model. 
In our quest to understand the geometrical properties of interfaces, we made a significant step forward by choosing Dobrushin boundary conditions to achieve a stationary state. 
Interestingly, this led us to the same conclusion as reported in \cite{holmes}, even before we became aware of this reference.

This study also confirms the singular nature of the voter model and its dual character, intermediate between the zero-temperature and critical Ising-Glauber models. 
Figuratively speaking, the voter model is cold inside and hot outside, while the critical Ising-Glauber model is equally warm throughout.

\begin{figure}[!ht]
\begin{center}
\includegraphics[angle=0,width=0.5\linewidth,clip=true]{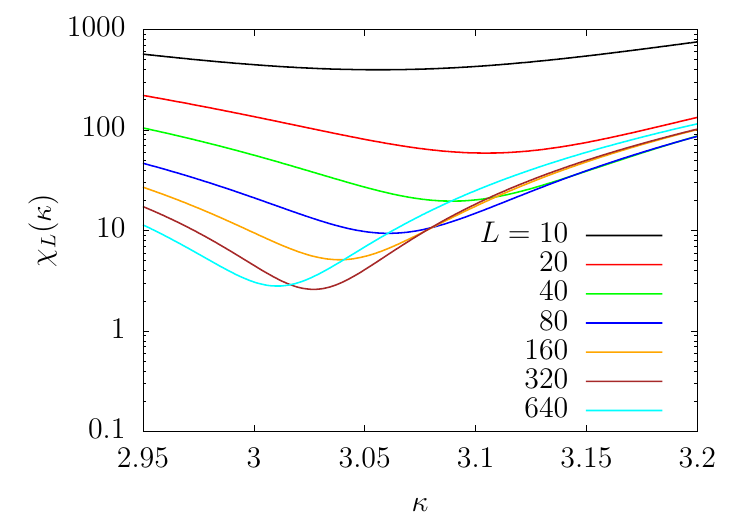}
\hskip -0.2cm
\includegraphics[angle=0,width=0.5\linewidth,clip=true]{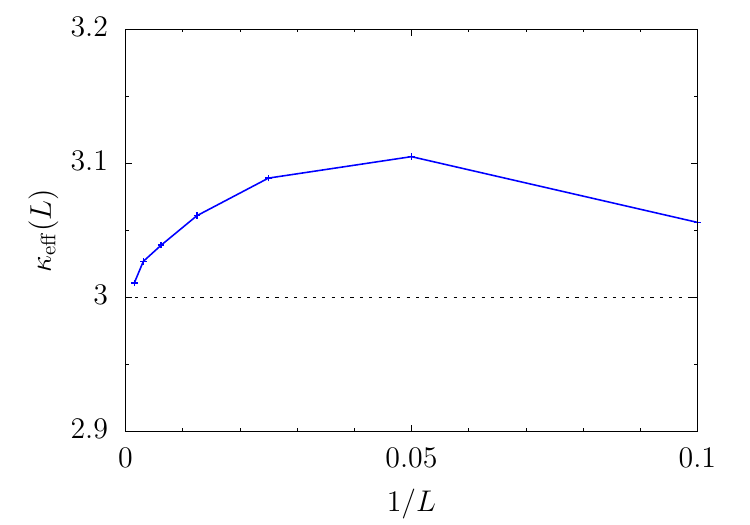}
\end{center}
\caption{Left panel: 
$\chi_L(\kappa)$ against $\kappa$ for the critical majority vote model.
Right panel: $\kappa_{\mathrm{eff}}(L)$ against $1/L$.}
\label{fig:MV}
\end{figure}

Additionally, this study demonstrates that the Ising universality class, which encompasses both static and dynamic properties, extends to include the geometrical properties of interfaces. 
For example, all conclusions drawn in the present work for the critical Ising-Glauber model are also applicable to the critical majority vote model, even though the latter is inherently a nonequilibrium model.
This is illustrated by figure \ref{fig:MV}, which depicts the result of the same analysis as that presented 
in section \ref{sec:leftproba}.
The critical value of the parameter $\delta$, as defined in  (\ref{eq:mvm}), is numerically determined to be $\delta_c\approx 0.84961$.
The same property (i.e., convergence to $\kappa=3$) holds for the case of the model located at the end of the critical line, where $(\gam_{\mathrm{int}}=1, \gam_{\mathrm{bulk}}\approx 0.41795)$.
Using the same figure of speech as previously, this model is warm inside and cold outside.
Yet, it belongs to the Ising universality class, too.

\ack
It is a pleasure to acknowledge interesting conversations with J M Luck.
\section*{References}
\bibliography{paper.bib}

\end{document}